\documentclass[preprint,3p,times,twocolumn,10p]{elsarticle}  

\usepackage{lineno,hyperref, amsmath, graphicx, multirow}
\usepackage{algorithm, algorithmic}

\begin{document}

\begin{frontmatter}

\title{End-to-End Learned Block-Based Image Compression with Block-Level Masked Convolutions and Asymptotic Closed Loop Training}





\author[focal]{Fatih~Kamisli}
\ead{kamisli@metu.edu.tr}

\address[focal]{Department of Electrical and Electronics Engineering, Middle East Technical University, Turkey}

\begin{abstract}
Learned image compression research has achieved state-of-the-art compression performance with auto-encoder based neural network architectures, where the image is mapped via convolutional neural networks (CNN) into a latent representation that is quantized and processed again with CNN to obtain the reconstructed image. CNN operate on entire input images. On the other hand, traditional state-of-the-art image and video compression methods process images with a block-by-block processing approach for various reasons. Very recently, work on learned image compression with block based approaches have also appeared, which use the auto-encoder architecture on large blocks of the input image and introduce additional neural networks that perform intra/spatial prediction and deblocking/post-processing functions. This paper explores an alternative learned block-based image compression approach in which neither an explicit intra prediction neural network nor an explicit deblocking neural network is used. A single auto-encoder neural network with block-level masked convolutions is used and the block size is much smaller (8x8). By using block-level masked convolutions, each block is processed using reconstructed neighboring left and upper blocks both at the encoder and decoder. Hence, the mutual information between adjacent blocks is exploited during compression and each block is reconstructed using neighboring blocks, resolving the need for explicit intra prediction and deblocking neural networks. Since the explored system is a closed loop system, a special optimization procedure, the asymptotic closed loop design, is used with standard stochastic gradient descent based training. The experimental results indicate competitive image compression performance.
\end{abstract}


\end{frontmatter} 


\newcommand{\xh}{\hat{x}}
\newcommand{\yh}{\hat{y}}
\newcommand{\zh}{\hat{z}}
\newcommand{\xt}{\tilde{x}}
\newcommand{\yt}{\tilde{y}}
\newcommand{\zt}{\tilde{z}}
\newcommand{\tu}{\ensuremath{\tilde{u}(i,j)}}
\newcommand{\tuq}[2]{\ensuremath{\tilde{u}(#1,#2)}}
\newcommand{\tuqh}[2]{\ensuremath{\hat{\tilde{u}}(#1,#2)}}

\section{Introduction}  
\label{sec:intro}
The\footnote{Codes are shared at \href{https://github.com/metu-kamisli/Learned-block-based-image-compression}{https://github.com/metu-kamisli/Learned-block-based-image-compression}} success of deep neural networks (DNN) in computer vision applications has attracted great attention \cite{guo2016deep} and many problems in image processing  have been revisited with the tools that DNN provide \cite{tian2018deep,litjens2017survey,wang2020deep,minaee2021image,ma2019image}. One such problem is image compression. Artificial neural network (ANN) based image compression approaches flourished quickly and their compression performance have caught up in recent years with that of traditional image and video compression systems \cite{hu2021learning}.   

Early traditional lossy image compression methods like JPEG \cite{JPEG} and JPEG2000 \cite{jp2k} use a linear decorrelating  transform to map pixel values to a transform domain, apply scalar quantization and process the quantized transform coefficients with the inverse transform to reconstruct the compressed image \cite{goyal2001}. More recent image compression systems, such as Webp, BPG and AVIF, are based on intra-frame compression tools of video compression standards, such as HEVC \cite{hevcintra} or AV1 \cite{av1}. These systems operate with a block-by-block processing approach and use spatial prediction prior to the linear decorrelating transform and are designed as multiple-mode compression systems. In particular, there are multiple spatial prediction modes, multiple linear transforms and multiple block sizes along with other multiple-mode tools. These multiple-mode tools allow the compression algorithm to adapt to varying spatial characteristics in images and improve the compression performance. Note that although the processing in each spatial prediction mode or transform is linear, the collection of all multiple-mode tools and a mode decision algorithm at the encoder constitute overall a nonlinear compression system.

ANN based compression systems, on the other hand, typically do not have multiple modes of operation but are inherently nonlinear (or adapt to image content) as they use ANN. They mostly use an auto-encoder based architecture which uses convolutional neural networks (CNN) as nonlinear analysis and synthesis transforms \cite{balle2016end,balle2018variational}. The nonlinear analysis transform maps pixels values to a latent domain, where scalar quantization is performed. The quantized latent domain variables are then mapped back with the nonlinear synthesis transform to the pixel domain to reconstruct the compressed image. These nonlinear transforms are learned by minimizing a rate-distortion cost over a large set of training images. To have efficient entropy coding, the  probability distribution of the quantized latent domain variables is also learned with an ANN along with the analysis and synthesis transforms. The entire compression system can be trained end-to-end to minimize a rate-distortion cost function consisting of any differential distortion metric, such mean-squared-error (MSE) or multi-scale structural similarity index measure (MS-SSIM) \cite{ssim, ms-ssim}, and the estimated entropy (rate) of the quantized latent variables via a differential relaxation method to avoid the non-differential quantization operation \cite{balle2016end,balle2018variational}.  

Most ANN based compression systems use CNN to form the nonlinear analysis and synthesis transforms (and entropy models). CNN operate on the entire input image \cite{GU2018354}. On the other hand, traditional state-of-the-art image and video compression methods use a block-by-block processing approach for various reasons \cite{LEGALL1992129, luthra264, HEVC}. The  primary reason is that the motion-compensated inter-frame coding method that dominates the video compression system design is inherently a block-based system and other aspects of video compression systems are also designed to operate block-by-block. In addition, switching between temporal and spatial (i.e inter and intra) prediction on a block-by-block basis is easy to achieve and improves compression performance. 

Very recently, few papers on learned image compression with block based approaches have also appeared \cite{blk2021wu, blk2021yuan, blk2021zhao}. They split the input image into large blocks (256x256 \cite{blk2021wu}, 128x128 \cite{blk2021wu, blk2021zhao}, 64x64 \cite{blk2021wu, blk2021yuan}, 32x32 \cite{blk2021yuan}) and use the described auto-encoder architecture \cite{minnen2018joint} on each block but introduce additional ANN that perform spatial/intra prediction \cite{blk2021wu, blk2021yuan} (to exploit mutual information between adjacent blocks) and post-processing/deblocking \cite{blk2021wu, blk2021yuan, blk2021zhao} functions (to combine the reconstructed image blocks smoothly with each other). 

This paper explores an alternative learned block-based image compression approach in which neither an explicit intra/spatial prediction ANN nor a deblocking ANN is used. A single auto-encoder ANN with block-level masked convolutions is used and the block size is much smaller (8x8). By using block-level masked convolutions in the auto-encoder, each block is processed using neighboring left and upper blocks both at the encoder and decoder. Hence, the mutual information between adjacent blocks is exploited during compression and there is no need for an explicit intra/spatial prediction ANN. At the decoder, each block is reconstructed using neighboring blocks and there is no need for an explicit post-processing/deblocking ANN. 

One advantage of the approach explored in this paper, relative to other block based learned compression approaches in the literature, is that it has a simpler and cleaner architecture consisting of one auto-encoder based ANN. Another potential advantage is that it can be easily extended to video (i.e. inter-frame) compression \cite{LEGALL1992129} by incorporating motion-compensated reference blocks instead of the spatial neighbor blocks.
Hence, as a byproduct, switching between inter-frame and intra-frame compression can be achieved simply and the small block-size of 8x8 is likely to allow for better compression-efficiency relative to switching between the compression methods in approaches that use larger block sizes.


The experimental results indicate that the explored learned block-based image compression approach in this paper provides competitive compression performance. Compared to HEVC (BPG) intra coding, the performance is slightly inferior at low bitrates but superior at higher bitrates. Compared to state-of-the-art learned compression approaches which do not operate on a block-by-block basis, the performance is slightly inferior or similar at low bitrates but catches up or is superior at higher bitrates. More details are provided in Section \ref{sec:exp}.

The remainder of the paper is organized as follows. Section \ref{sec:relwork} reviews related work. Section \ref{sec:prop} presents the explored learned block-based compression approach. The end-to-end training of the explored system, including  the Asymptotic Closed Loop design and other details are discussed in Section \ref{sec:tr}. Section \ref{sec:exp} presents experimental results and comparisons with other work. Finally, Section \ref{sec:conc} concludes the paper.

\section{Related Work}
\label{sec:relwork}

\subsection{End-to-end Learned Image Compression Review}
\label{sec:eework}
While research on learned image compression can be traced back many years \cite{jiang1999image, toderici2015variable} the recent surge of interest started with the works of Balle et. al \cite{balle2016end, balle2018variational, minnen2018joint}, and since then, many researchers used the compression architecture in \cite{minnen2018joint} as starting point and extended it in many ways.   

\begin{figure}[tb]
\begin{minipage}{0.45\linewidth}
\includegraphics[trim=48 710 400 46,clip,width=\linewidth]{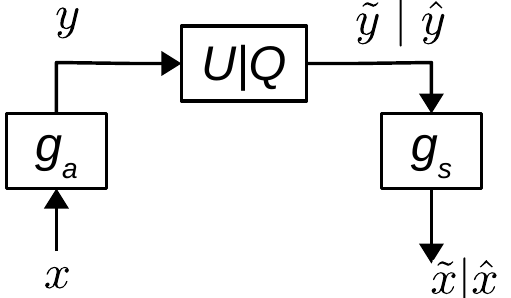}
\centerline{{\footnotesize (a) Baseline model \cite{balle2016end}}}
\end{minipage}
\hfill
\begin{minipage}{0.45\linewidth}
\includegraphics[trim=220 635 220 46,clip,width=\linewidth]{./figs/basemodel.pdf}
\centerline{{\footnotesize (b) Hyper-prior based model \cite{balle2018variational}}}
\end{minipage}
\hfill
\center
\begin{minipage}{0.50\linewidth}
\includegraphics[trim=400 615 32 40,clip,width=\linewidth]{./figs/basemodel.pdf}
\centerline{{\footnotesize (c) Joint auto-regressive and hyper-prior based model \cite{minnen2018joint}}}
\end{minipage}
\caption{Operational diagrams of learned compression models}
\label{fig:ballemodels}
\end{figure}

The baseline ANN based compression model in \cite{balle2016end} is shown in Figure \ref{fig:ballemodels} a). The image $x$ is processed with the nonlinear analysis function $g_a(.)$, which consists of 3 successive layers of CNN with downsampling factors 2 (or 4) and nonlinear generalized divisive normalization (GDN) operations, to obtain the latent representation $y$. The latent $y$ is quantized and the quantized latent $\hat{y}$ is lossless coded with an arithmetic coder. (The $U|Q$ symbol denotes quantization to nearest integer in compression/inference and addition of uniform noise during training.) The decoder decodes the bitstream to obtain $\hat{y}$ and processes it with the nonlinear synthesis function $g_s(.)$, which also consists of 3 successive layers of CNN with an upsampling factor of 2 (or) and a nonlinear inverse of generalized divisive normalization (IGDN) operation, to reconstruct the compressed image $\hat{x}$.

The analysis and synthesis transforms as well as the probability distribution of the quantized latent $\hat{y}$ are learned by minimizing a rate-distortion cost $L$ over a set of training images:
\begin{align}
  L & = R(\yh) + \lambda D(x,\xh) \\
    & = E[-log_2 p_{\yh}(\yh)] + \lambda \cdot E[d(x,\xh)]
  \label{eq:rd}
\end{align} 
Here, $\lambda$ is a Lagrangian multiplier that controls the trade-off between rate and distortion and $d(.,.)$ is any differentiable distortion metric such as mean-squared-error (MSE) or multi-scale structural similarity index measure (MS-SSIM) \cite{ms-ssim}. The rate term $E[-log_2 p_{\yh}(\yh)]$ is obtained from the probability distribution $p_{\yh}(\yh)$ of $\yh$ assuming it is factorized over the distribution of elements $\yh_i$ as 
\begin{align}
  p_{\yh}(\yh) = \prod_i p_{\yh_i}(\yh_i)
\end{align} 
with the elements $\yh_i$ in the same channel being identically distributed \cite{balle2016end, balle2018variational}.

Since quantization operation has zero derivatives almost everywhere, the rate term is relaxed with $E[-log_2 p_{\yt}(\yt)]$ where the probability density function $p_{\yt}(\yt)$ is similarly factorized over the densities of the elements $\yt_i$, which are obtained by adding noise $u_i$ uniformly distributed in $[-\frac{1}{2}, \frac{1}{2}]$ to the latent variables $y_i$: 
\begin{align}
  \yt_i = y_i + u_i 
  \label{eq:adns}
\end{align}
This relaxation allows to use $p_{\yt_i}(\yt_i)$ as a proxy for $p_{\yh_i}(\yh_i)$ during training since for integers $k$, $p_{\yt_i}(k)=p_{\yh_i}(k)$ i.e. the probability mass and density functions are tied together. The density values $p_{\yt_i}(\yt_i)$ are obtained from an ANN that models/learns the cumulative distribution function (CDF) $F_{y_i}(.)$ of $y_i$ based on Equation (\ref{eq:adns}) as \cite{balle2016end, balle2018variational}:
\begin{align}
  p_{\yt_i}(\yt_i) = F_{y_i}(\yt_i+\frac{1}{2}) - F_{y_i}(\yt_i-\frac{1}{2})
  \label{eq:cdf_cdf}
\end{align}

With this relaxation, the entire compression system can be trained end-to-end to determine the parameters of ANN forming $g_a(.)$, $g_s(.)$ and $F_y(.)$ by minimizing the rate-distortion cost in Equation (\ref{eq:rd}) over a set of training images.

The baseline ANN based compression model was improved in \cite{balle2018variational} by transmitting a learned hyper-prior $\zh$ (similar to side or forward information in compression theory) that will allow the decoder to decode the latent $\yh$ with a probability model that is spatially adapted to the coded image (see Figure \ref{fig:ballemodels} b). The hyper-prior $z$ is generated from the latent $y$ using a hyper-encoder $h_a(.)$ ANN, then quantized and lossless coded. The decoder decodes the bitstream of the hyper-prior and then processes it with the hyper-decoder $h_s(.)$ ANN to generate per-variable mean and standard deviations ($\mu_i, \sigma_i$) for the conditional probability distribution $p_{\yh|\zh}(\yh|\zh)$ of the latent $\yh$ conditioned on the hyper-prior $\zh$.

The rate-distortion cost to be minimized becomes 
\begin{align}
  L & = R(\zh) + R(\yh) + \lambda D(x,\xh) \\
    & = E[-log_2 p_{\zh}(\zh)] + E[-log_2 p_{\yh|\zh}(\yh|\zh)] + \lambda \cdot E[d(x,\xh)] 
  \label{eq:rd_h}
\end{align} 
where the rate of the hyper-latent $E[-log_2 p_{\zh}(\zh)]$ is now relaxed and modeled as the latent in the base model. 
The rate of the latent $E[-log_2 p_{\yh|\zh}(\yh|\zh)]$ is now relaxed via $E[-log_2 p_{\yt|\zt}(\yt|\zt)]$ during training where the conditional density $p_{\yt|\zt}(\yt|\zt)$ is factorized over the densities $p_{\yt_i|\zt}(\yt_i|\zt)$ of the elements $\yt_i$, which are obtained by adding noise uniformly distributed in $[-\frac{1}{2}, \frac{1}{2}]$ to the latent variables $y_i$. The conditional density values $p_{\yt_i|\zt}(\yt_i|\zt)$ are obtained using conditional CDFs $F_{y_i|\zt}(.)$, similar to Equation (\ref{eq:cdf_cdf}), and $F_{y_i|\zt}(.)$ are modeled with the Gaussian CDF whose mean and variance is given by $h_s(.)$, i.e.  
\begin{align}
  p_{y_i|\zt}(y_i|\zt) \sim \mathcal{N} (\mu_i, \sigma_i).
\end{align}

The hyper-prior based compression model was also improved in \cite{minnen2018joint} by forming the conditional density of the latent variables not only from the hyper-prior but also from the causal context $\yt_{<i}$ of each latent variable:
\begin{align}
  p_{y_i|\zt, \yt_{<i}}(y_i|\zt, \yt_{<i}) \sim \mathcal{N} (\mu_i, \sigma_i)
  \label{eq:jnt}
\end{align}
This joint (auto-regressive and hierarchical) priors model (see Figure \ref{fig:ballemodels} c) further improves the compression performance, however, due to the spatial dependency introduced by the causal context $\yt_{<i}$ the decoding process can not by fully parallelized anymore \cite{minnen2018joint}.

Many researchers have used the joint (auto-regressive and hierarchical) priors model compression architecture in \cite{minnen2018joint} and modified some of its components in several ways to obtain better compression performance \cite{cheng2020learned, akbari2020generalized, yilmaz2021self}. For example, Cheng et al. \cite{cheng2020learned} use attention modules and residual connections in the auto encoder architecture and replace the Gaussian distribution model in Equation (\ref{eq:jnt}) with a Gaussian mixture model. Akbari et al. \cite{akbari2020generalized} replace the standard CNN layers with generalized octave convolution layers to work with multi-frequency feature maps. Yilmaz et al. \cite{yilmaz2021self} replace the CNN and GDN layers in the auto encoder with self-organized operational neural network layers.              

\subsection{Block Based Learned Image Compression}
\label{sec:bbwork}
Recently, papers on learned image compression with block based approaches have also appeared \cite{blk2021wu, blk2021yuan, blk2021zhao}. They split the input image into large blocks (256x256 \cite{blk2021wu}, 128x128 \cite{blk2021wu, blk2021zhao}, 64x64 \cite{blk2021wu, blk2021yuan}, 32x32 \cite{blk2021yuan}) and use the above summarized joint model architecture in \cite{minnen2018joint} on each block but introduce additional ANN that perform spatial/intra prediction and post-processing/deblocking functions. The spatial/intra prediction ANN are used to exploit the correlation from the previously reconstructed left and upper blocks to improve the compression of the current block \cite{blk2021wu, blk2021yuan}. The post-processing/deblocking ANN are used to remove blocking artifacts that occur at the block boundaries mostly due to independent processing of the blocks \cite{blk2021wu, blk2021yuan, blk2021zhao}.

\section{Explored Block Based Image Compression Approach}  
\label{sec:prop}

This paper explores an alternative learned block-based image compression approach in which neither an explicit intra/spatial prediction ANN nor a deblocking ANN is used. A single auto-encoder based ANN with block-level masked convolutions is used and the block size is much smaller (8x8). By using block-level masked convolutions in the auto-encoder ANN, each block is processed using neighboring left and upper blocks both at the encoder and decoder. Hence, the mutual information between adjacent blocks is exploited during compression and there is no need for an explicit intra/spatial prediction ANN. At the decoder, each block is reconstructed using neighboring blocks and there is no need for an explicit post-processing/deblocking ANN.

The details of the system architecture and its operation during compression/inference is discussed next. Section \ref{sec:tr} discusses the details of training the system.

\subsection{ANN Architecture and Operation During Compression/Inference}
\label{ssec:inf}
\begin{figure*}[tbh]
\begin{minipage}{0.55\linewidth}
\centering
\includegraphics[trim=45 550 200 22,clip,width=\linewidth]{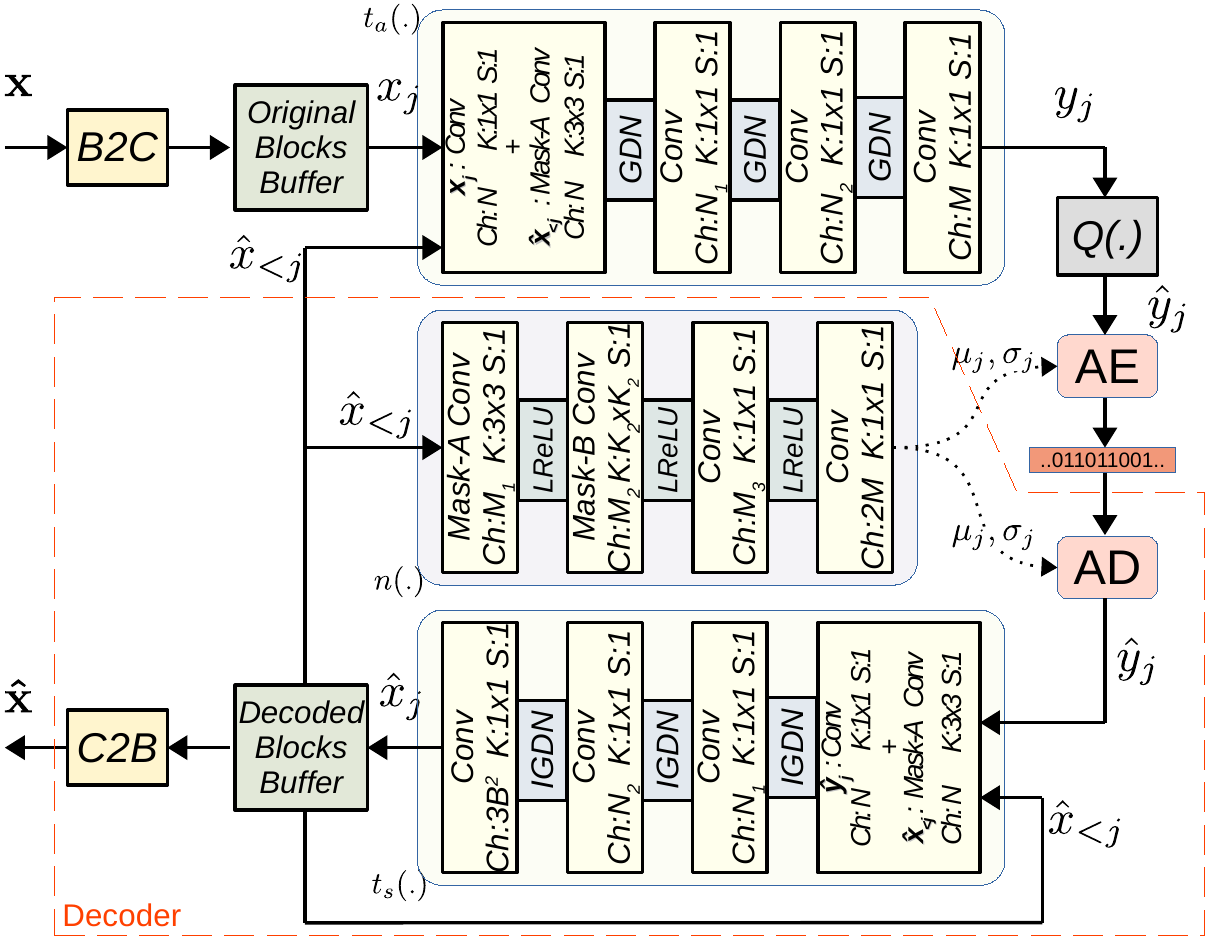}
\caption{Operational diagram for compression/inference. The encoder contains all subsystems (except AD) in the figure while the decoder contains the subsystems inside the red dashed lines. Both the encoder and decoder operate closed loop in a block-by-block manner.}
\label{fig:prop_inf} 
\end{minipage}
\hfill
\begin{minipage}{0.41\linewidth}
\centering
\includegraphics[trim=19 675 395 28,clip,width=0.80\linewidth]{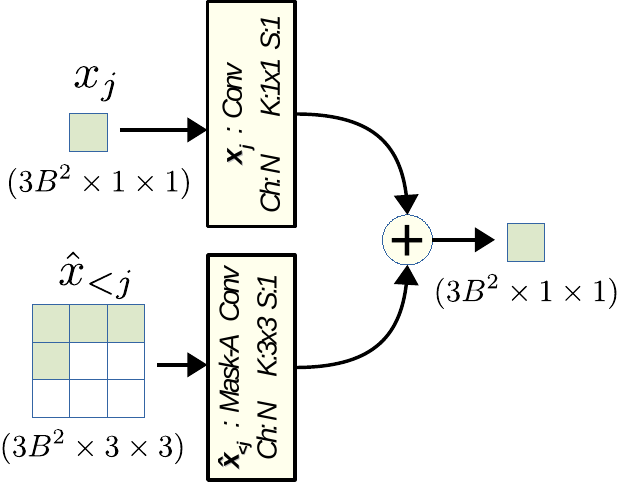}
\caption{Fusion of original block $x_j$ and reconstructed blocks in the causal neighborhood $\hat{x}_{<j}$ at the encoder with CNN in the first layer of $t_a(.)$. Masked convolution \cite{van2016conditional} is used to use only the causal neighborhood of, i.e. left and upper, reconstructed blocks.} 
\label{fig:mconv_fusion}
\end{minipage}
\end{figure*}

The operational diagram of the system for compression/inference is shown in Figure \ref{fig:prop_inf}. First, the image $\mathbf{x}$ is processed with the $B2C$ (block to channel) subsystem, which scans the pixels inside a $3$x$B$x$B$ block and arranges them along the channel dimension, i.e. an (RGB) image of size $3$x$H$x$W$ is converted to a tensor of size $3B^2$x$\frac{H}{B}$x$\frac{W}{B}$. This conversion allows to use the CNN routines in common deep learning frameworks for block-level convolutions. 

%

The system uses a block-by-block processing (in raster scan order) approach as traditional image and video compression systems. Hence, one block, shown with $x_j$ in Figure \ref{fig:prop_inf}, is processed through the entire system and its reconstruction, denoted with $\hat{x}_j$, is stored in the decoded blocks buffer to be used for the processing of the next blocks. 

At the encoder, the processing is performed as follows. In the first step, the CNN processing result of an original block $x_j$ (of size $3B^2$x$1$x$1$) is added to the CNN processing result of its previously reconstructed causal neighborhood of blocks $\hat{x}_{<j}$ (of size $3B^2$x$3$x$3$) as shown in Figure \ref{fig:mconv_fusion}. The result is then further processed with GDN (Generalized Divisive Normalization \cite{balle2015density, balle2016end}) activation functions and CNN layers to produce the latent variables block $y_j$. This processing at the encoder is denoted with $t_a(.)$ in Figure \ref{fig:prop_inf} and can be seen as a nonlinear predictive analysis transform, playing the role of intra prediction and transforming the prediction residual in traditional image compression. 
\begin{align}
  y_j = t_a(x_j, \hat{x}_{<j})
\end{align}

The latent block is quantized (by rounding each element to the nearest integer) to produce the quantized latent block $\hat{y}_j$, which is lossless compressed into a bitstream by an arithmetic encoder (AE) as shown in Figure \ref{fig:prop_inf}. The probability distribution that the AE uses is determined by the parameters $\mu_j$ and $\sigma_j$ that are produced by an ANN, denoted by $n(.)$ in Figure \ref{fig:prop_inf}, which processes the previously reconstructed causal neighborhood of blocks $\hat{x}_{<j}$. This ANN consists of CNN with masked convolutions (to capture only the causal context) or regular convolutions (with kernel size 1x1) and LeakyReLU activation functions. Note that this ANN based subsystem $n(.)$ is common to both the encoder and decoder as they need to generate the same probability distribution $p_{\hat{y}_j|\hat{x}_{<j}}(\hat{y}_j|\hat{x}_{<j})$ that both the AE and the arithmetic decoder (AD) must use. These processing steps can be summarized as follows:
\begin{align}
  \hat{y}_j = Q(y_j) \\
  (\mu_j, \sigma_j) = n(\hat{x}_{<j}) \\
  p_{\hat{y}_j|\hat{x}_{<j}}(\hat{y}_j|\hat{x}_{<j}) \sim (\mu_j, \sigma_j).     
\end{align}

The quantized latent variable block $\hat{y}_j$ is processed together with previously reconstructed causal neighborhood of blocks $\hat{x}_{<j}$ via the reconstruction ANN, denoted with $t_s(.)$ in Figure \ref{fig:prop_inf}, to obtain the reconstruction block $\hat{x}_j$.
\begin{align}
  \hat{x}_j = t_s(\hat{y}_j, \hat{x}_{<j})
\end{align}
In the first step of the reconstruction ANN, the CNN processing result of the quantized latent block $\hat{y}_j$ (of size $M$x$1$x$1$) is added to the CNN processing result of the previously reconstructed causal neighborhood of blocks $\hat{x}_{<j}$ (of size $3B^2$x$3$x$3$), similar to the first step of processing at the encoder. The result is then further processed with IGDN (Inverse Generalized Divisive Normalization \cite{balle2015density, balle2016end}) and CNN layers to produce the  reconstructed block $\hat{x}_j$. This reconstruction process at the decoder, denoted with $t_s(.)$ in Figure \ref{fig:prop_inf}, can be seen as a nonlinear predictive synthesis transform, playing the role of the following operations in traditional image compression: inverse transforming the quantized transform coefficients, performing intra prediction and adding their results. The reconstructed block $\hat{x}_j$ is stored in reconstructed blocks buffer and the compression/decompression of the next block can start.

The decoder works in a similar manner with block-by-block processing and consists of the sub-systems in the red dashed box in Figure \ref{fig:prop_inf}. First, the ANN $n(.)$ processes causal context $\hat{x}_{<j}$ to obtain probability distribution parameters $\mu_j$ and $\sigma_j$ that are used by the AD to obtain quantized latent block $\hat{y}_{j}$. Then $\hat{y}_{j}$ and causal context $\hat{x}_{<j}$ are processed jointly by ANN $t_s(.)$ to reconstruct the block $\hat{x}_{j}$.

Finally, note that the block-by-block processing prevents CNN computations to be performed on the entire images or feature maps, i.e. one block has to be processed through the entire system including all CNN and then the next block processing can start. Such computations are also present in some other learning based approaches \cite{minnen2018joint} and may not fully capitalize on the computational resources of standard GPU hardware or software. More details are discussed in Section \ref{ssec:companal}.

\section{End-to-End Training with Asymptotic Closed Loop Design} 
\label{sec:tr}
\subsection{Asymptotic Closed Loop (ACL) Optimization}
\label{ssec:acl}
The explored block based learned image compression system is a closed loop system. In other words, the compression result $\hat{x}_j$ of one block is used for the compression of the next blocks. This poses a fundamental problem for the end-to-end training of the system since initially the compression system with randomly initialized ANN weights cannot produce proper reconstructed blocks $\hat{x}_{<j}$ using which the system can be trained. 

A naive approach would be to use the original (uncompressed) blocks in the causal neighborhood $x_{<j}$ instead of the reconstructed blocks $\hat{x}_{<j}$ during training. This is termed the open-loop optimization procedure in \cite{acl2001} and produces inferior compression performance when such trained system is used in the actual compression/inference, which operates closed loop. In particular, the compression/inference performance is catastrophic at low bit-rates where the reconstructed blocks and original blocks differ significantly.

We adopt the Asymptotic Closed Loop (ACL) design procedure from \cite{acl2001}, which was proposed for predictive vector quantizer design in signal compression and was also applied to predictor design in video compression \cite{vishwanath2021effective}. The ACL procedure provides stable open-loop training while ultimately optimizing the compression system for closed loop operation. 

\begin{figure*}[bt]
\begin{minipage}{0.55\linewidth}
\centering
\includegraphics[trim=41 190 135 385,clip,width=1.0\linewidth]{./figs/our_nn_arch.pdf}
\caption{Operational diagram during training. System operates open loop.}
\label{fig:prop_tr}
\end{minipage}
\hfill
\begin{minipage}{0.40\linewidth}
\centering
\includegraphics[trim=115 250 120 265,clip,width=1.0\linewidth]{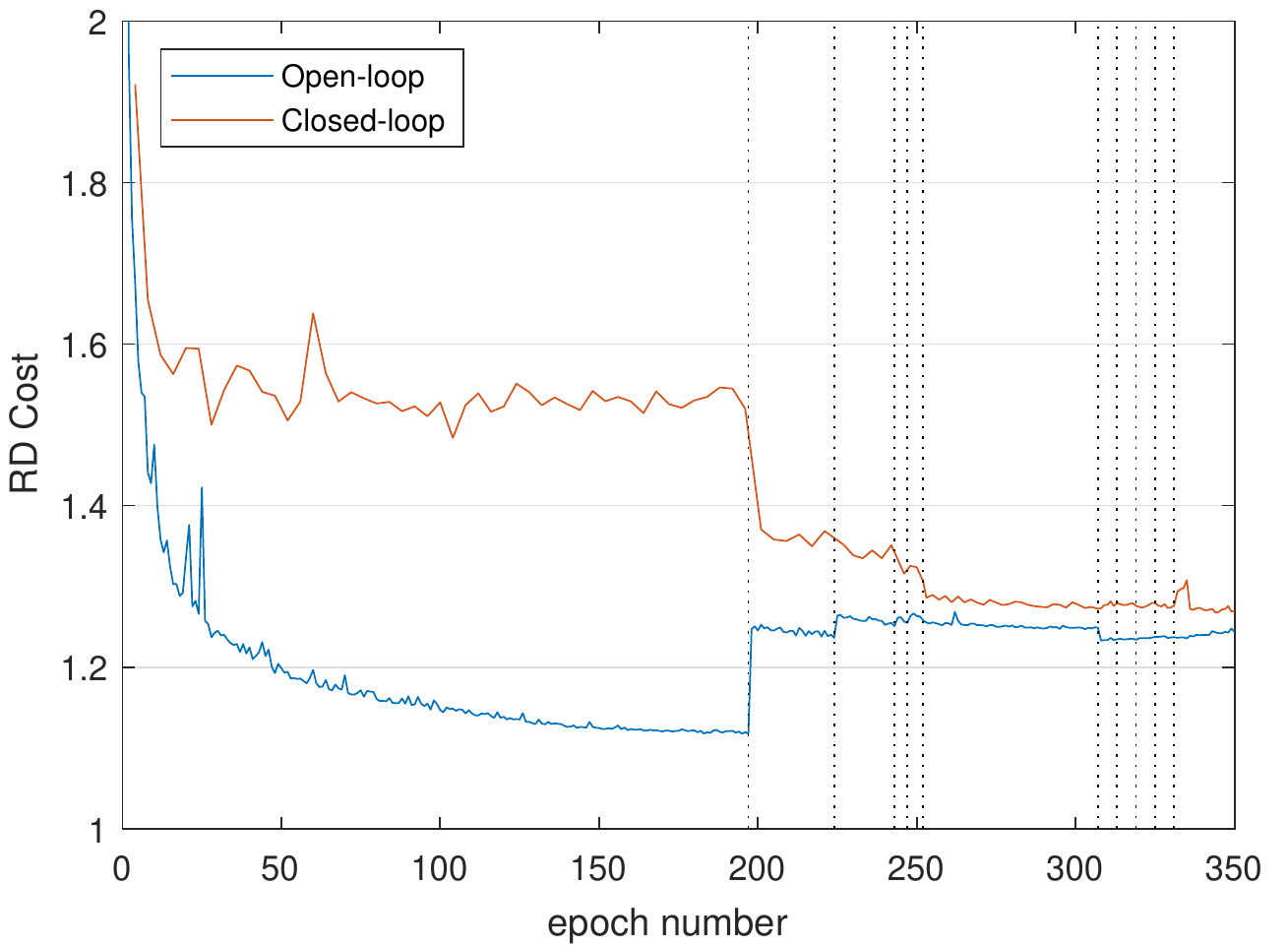} 
\caption{Sample validation set RD Cost curves of open loop and closed loop operations over SGD epochs}
\label{fig:acl_itrs}
\end{minipage}
\end{figure*}

The ACL design algorithm proceeds in iterations where in each iteration the compression system is trained open loop (with standard stochastic gradient descent) until convergence. The operational diagram of the compression system in open loop training is shown in Figure \ref{fig:prop_tr}. In the $k^{th}$ ACL iteration, the reconstructions $\mathbf{\hat{x}}^{(k)}$ that the system uses in the training are not taken from the output of the system but from a secondary training set, which was produced with the converged system at the end of the $(k-1)^{th}$ ACL iteration. In the initial iteration ($k=0$), the system uses the original images also for the reconstructions, i.e. $\mathbf{\hat{x}}^{(0)} = \mathbf{x}$. 


Another difference between the open loop operation of the system in the ACL iterations and its closed loop operation during actual compression/inference is that the block-by-block processing is not necessary in the open loop operation. In other words, entire images can be processed through the system at once, which speeds up training. Hence, the original blocks buffer and decoded blocks buffer in Figure \ref{fig:prop_inf} are not present in Figure \ref{fig:prop_tr}. A summary of the used ACL design algorithm is given in Algorithm \ref{alg:acl}. 

With every ACL iteration, the system is trained with reconstructions $\mathbf{\hat{x}}^{(k)}$ that are closer to its produced reconstructions $\mathbf{\tilde{x}}$ and the closed loop compression performance improves. We observe convergence after several ACL iterations. Sample curves of the validation RD costs of the open loop and closed loop operations of the system are shown in Figure \ref{fig:acl_itrs}. The curves are plotted for many SGD training epochs, which cover several ACL iterations. The end of each ACL iteration is shown with vertical dashed lines. It can be seen that in the initial ($k=0$) iteration, there is a big gap between the open and closed loop RD costs, which reduces with following ACL iterations.

\begin{algorithm}[t]
  \small
    \caption{Asymptotic Closed Loop (ACL) training}\label{your_label}
    \begin{algorithmic}
        \STATE Define training and validation images sets : $\{\mathbf{x}\}, \{\mathbf{x_{VAL}}\}$  
        \STATE Initialize reconstructed images set : $\{\mathbf{\hat{x}}^{(0)}\} = \{\mathbf{x}\}$
        \STATE Initialize compression system parameters : $\Phi^{(0)}$ $\leftarrow$ random
        \FOR{ACL iteration k $=0$ to $\infty$}
            \STATE $\mathbf{1.}$ Train compression system open loop with SGD until convergence : $\Phi^{(k+1)} \leftarrow train(\Phi^{(k)}, \{\mathbf{x}\}, \{\mathbf{\hat{x}}^{(k)}\})$
            \STATE $\mathbf{2.}$ Get compression systems closed loop validation cost : $RDCost^{(k+1)}_{VAL} \leftarrow \mathbf{\Phi_{CL}}^{(k+1)}(\{\mathbf{x_{VAL}}\})$
            \STATE $\mathbf{3.}$ Break if $RDCost^{}_{VAL}$ values converge
            \STATE $\mathbf{4.}$ Generate reconstructed images set for next iteration : $\{\mathbf{\hat{x}}^{(k+1)}\} \leftarrow \mathbf{\Phi_{OL}}^{(k+1)}(\{\mathbf{x}\}, \{\mathbf{\hat{x}}^{(k)}\})$
        \ENDFOR
    \end{algorithmic}
    \label{alg:acl}
\end{algorithm}


\subsection{Rate-Distortion Cost for SGD Optimization of System Parameters}
\label{ssec:tr}

In each ACL iteration, the compression system parameters are optimized to minimize the following rate-distortion (RD) cost $L^{(k)}$ over a set of training and reconstruction images, $\{\mathbf{x}\}$ and $\{\mathbf{x^{(k)}}\}$ respectively.
\begin{align}
  L^{(k)} & = R(\yh) + \lambda D(x,\xt) \\
         & = E[-log_2 p_{\yh|\xh^{(k)}}(\yh|\xh^{(k)})] + \lambda \cdot E[d(x,\xt)]
  \label{eq:prop_rd}
\end{align} 
Here, $\lambda$ is a Lagrangian multiplier that controls the trade-off between rate and distortion and $d(.,.)$ is a differentiable distortion metric such as mean-squared-error (MSE) or multi-scale structural similarity index measure (MS-SSIM). The rate $R(\yh)$ of the latent is obtained from the conditional distribution $p_{\yh|\xh^{(k)}}(\yh|\xh^{(k)})$ of the latent $\yh$ conditioned on the causal neighborhood of the reconstructions in $\xh^{(k)}$ and there is no hyper-latent transmission as in \cite{minnen2018joint} or the block-based approaches using its main architecture \cite{blk2021wu, blk2021yuan, blk2021zhao}.



Similar to \cite{balle2018variational, minnen2018joint}, the rate of the latent is relaxed via $E[-log_2 p_{\yt|\xh^{(k)}}(\yt|\xh^{(k)})]$ where the conditional density $p_{\yt|\xh^{(k)}}(\yt|\xh^{(k)})$ is factorized over the densities $p_{\yt_i|\xh^{(k)}}(\yt_i|\xh^{(k)})$ of the elements $\yt_i$, which are obtained by adding noise uniformly distributed in $[-\frac{1}{2}, \frac{1}{2}]$ to the latent variables $y_i$. The conditional density values $p_{\yt_i|\xh^{(k)}}(\yt_i|\xh^{(k)})$ are obtained using conditional CDFs $F_{y_i|\xh^{(k)}}(.)$, similar to Equation (\ref{eq:cdf_cdf}), and $F_{y_i|\xh^{(k)}}(.)$ are modeled with the Gaussian CDF whose mean and variance is given by the ANN $n(.)$ that processes the causal context of the reconstruction $\xh^{(k)}$ (see Figure \ref{fig:prop_tr}), i.e.  
\begin{align}
  p_{y_i|\xh^{(k)}}(y_i|\xh^{(k)}) \sim \mathcal{N} (\mu_i, \sigma_i).  
\end{align}

With this relaxation, the entire compression system can be trained end-to-end to determine the parameters of ANN forming $t_a(.)$, $t_s(.)$ and $n(.)$ by minimizing the rate-distortion cost in Equation (\ref{eq:prop_rd}) over a set of training images.

\subsection{Fine Tuning of System Parameters}
\label{ssec:ft}
Upon convergence of the ACL iterations, it is observed that the MSE of the reconstructions in successive iterations are very similar ( i.e. $d(\mathbf{x_{VAL}}, \mathbf{\xh_{VAL}^{(k)}}) \sim d(\mathbf{x_{VAL}}, \mathbf{\xh_{VAL}^{(k+1)}})$) but the closed loop validation RD cost
\[ RDCost^{(k+1)}_{VAL,CL} \leftarrow \mathbf{\Phi_{CL}}^{(k+1)}(\{\mathbf{x_{VAL}}\}) \]
can still be higher than the open loop validation RD cost
\[ RDCost^{(k+1)}_{VAL,OL} \leftarrow \mathbf{\Phi_{OL}}^{(k+1)}(\{\mathbf{x_{VAL}}\}, \{\mathbf{\hat{x}_{VAL}}^{(k)}\}) .\]

This small discrepancy hints that although at convergence of ACL iterations, successive iterations provide reconstructions with similar distortions, their statistics may still be slightly different since the closed loop operation does give slightly inferior RD cost than open loop operation. Thus, a fine-tuning stage was added where the reconstructions produced by the system ($\xt$ in Figure \ref{fig:prop_tr}) were fed back to the system in place of $\xh^{(k)}$ and the cost to be minimized was formed as average of the two RD costs from the first and second runs of the system as follows
\begin{align}
  L^{(k)} = & \frac{1}{2} ( E[-log_2 p_{\yt_0|\xh^{(k)}}(\yt_0|\xh^{(k)})] + \lambda \cdot E[d(x,\xt_0)] ) ~ + \nonumber \\
              & \frac{1}{2} ( E[-log_2 p_{\yt_1|\xt_0}(\yt_1|\xt_0)] + \lambda \cdot E[d(x,\xt_1)] ).
  \label{eq:prop_rdft}
\end{align} 
Here, $\xt_0$ and $\xt_1$ are the reconstructions and $\yt_0$ and $\yt_1$ are the latents from the first and second runs of the system. 

In summary, the system parameters were fine tuned for a few more ACL iterations following Algorithm \ref{alg:acl} except that in Step 1 the RD cost that was minimized is the one in Equation (\ref{eq:prop_rdft}). This improved the RD cost of the system in closed loop operation. In Figure \ref{fig:acl_itrs}, the fine-tuning ACL iterations start with the $5^{th}$ iteration after epoch 252.

\subsection{System Hyper Parameters}
\label{ssec:hp}
As in many learned image compression systems, a new set of system parameters is learned for each different rate-distortion (RD) trade-off by changing the $\lambda$ parameter in Equation (\ref{eq:prop_rd}) \cite{balle2016end, balle2018variational, minnen2018joint}. The explored system is trained for 8 different $\lambda$ values, corresponding to those in \cite{begaint2020compressai}, to cover a sufficiently large range of reconstructed picture qualities.

Two sets of hyper parameters were used for the explored system, which mainly differ in the number of channels used for the CNN in the system. One set of hyper parameters was used for the four smaller $\lambda$ values, and another set with larger learning capacity for the four larger $\lambda$ values where the system needs to reconstruct higher fidelity images. The hyper parameters are summarized in Table \ref{tb:hp}.

\begin{table}[h]
\begin{small}\begin{center}
\caption{Two sets of hyper parameters (HPS) to define the system}
\vspace{0.1cm}
{\renewcommand{\arraystretch}{1.2} 
\begin{tabular}{c|cc} 
  & HPS1 & HPS2 \\ \cline{1-3}
N & 768 & 1152 \\
M &  96 &  128 \\ \hline
$N_1$  &  \multicolumn{2}{c}{$\frac{7}{8}N$} \\  
$N_2$  &  \multicolumn{2}{c}{$\frac{6}{8}N$} \\ \hline
$M_1$  &  \multicolumn{2}{c}{$\frac{6}{4}N$} \\
$M_2$  &  \multicolumn{2}{c}{$\frac{5}{4}N$} \\
$M_3$  &  \multicolumn{2}{c}{$\frac{4}{4}N$} \\ \hline  
$K_2$x$K_2$ & 1x1 & 3x3 
\end{tabular}
}
\label{tb:hp} 
\end{center}\end{small}  
\end{table}

\subsection{Block Size}
\label{ssec:bsize}
A very important hyper parameter of the explored learned block based image compression system is the block size. The system supports a single block size, and a block size of 8x8 is used in the experimental results that are presented in the next section. This block size provides competitive overall compression results and reasonable computational/model complexity. A larger block size, such as 16x16, requires higher number of channels for the CNN to provide efficient compression and increases computational/model complexity significantly. A smaller block size, such as 4x4, fails to provide competitive compression results at lower bitrates since compression of large smooth regions becomes quite inefficient with 4x4 block size. In summary, the fixed block size of 8x8 is used for the main results presented in the next section, however some results and comments for 16x16 and 4x4 block size systems are also provided.

\section{Experimental Results}
\label{sec:exp}

\subsection{Training Set and Further Details}
\label{ssec:ts}
For training, the CLIC professional and mobile datasets \cite{CLIC2020} were used with randomly cropped patches of 256x256 and a batch size of 4 for each SGD iteration. The Adam optimizer \cite{adam} was used with an initial learning rate of $10^{-4}$ together with a learning rate scheduler that reduces the learning rate by a factor of $0.8$ (up to a minimum of $2$x$10^{-5}$) when the cost plateaus.     

The entire system and associated training algorithms were implemented with PyTorch and numerically stable Gaussian CDF implementations from \cite{begaint2020compressai} and the entropy coder/decoder of \cite{duda2014asymmetric} with \cite{begaint2020compressai}'s  implementation were utilized. The codes to repeat our results are shared at \cite{this-gh}.  

\subsection{Compression Results}
\label{ssec:res}
To test the compression performance of the explored learned block based image compression method, the Kodak \cite{kodak} and the Tecnick datasets \cite{tecnick} were used. The Kodak dataset contains a set of 24 uncompressed images with 512x768 resolution (and 4:4:4 RGB color sub sampling) commonly used to evaluate image compression methods. The Tecnick dataset contains 100 images of 1200x1200 (RGB 4:4:4) resolution. 

\begin{figure*}[tb]
\begin{minipage}{0.49\linewidth}
\includegraphics[trim=20 15 33 35, clip, width=\linewidth]{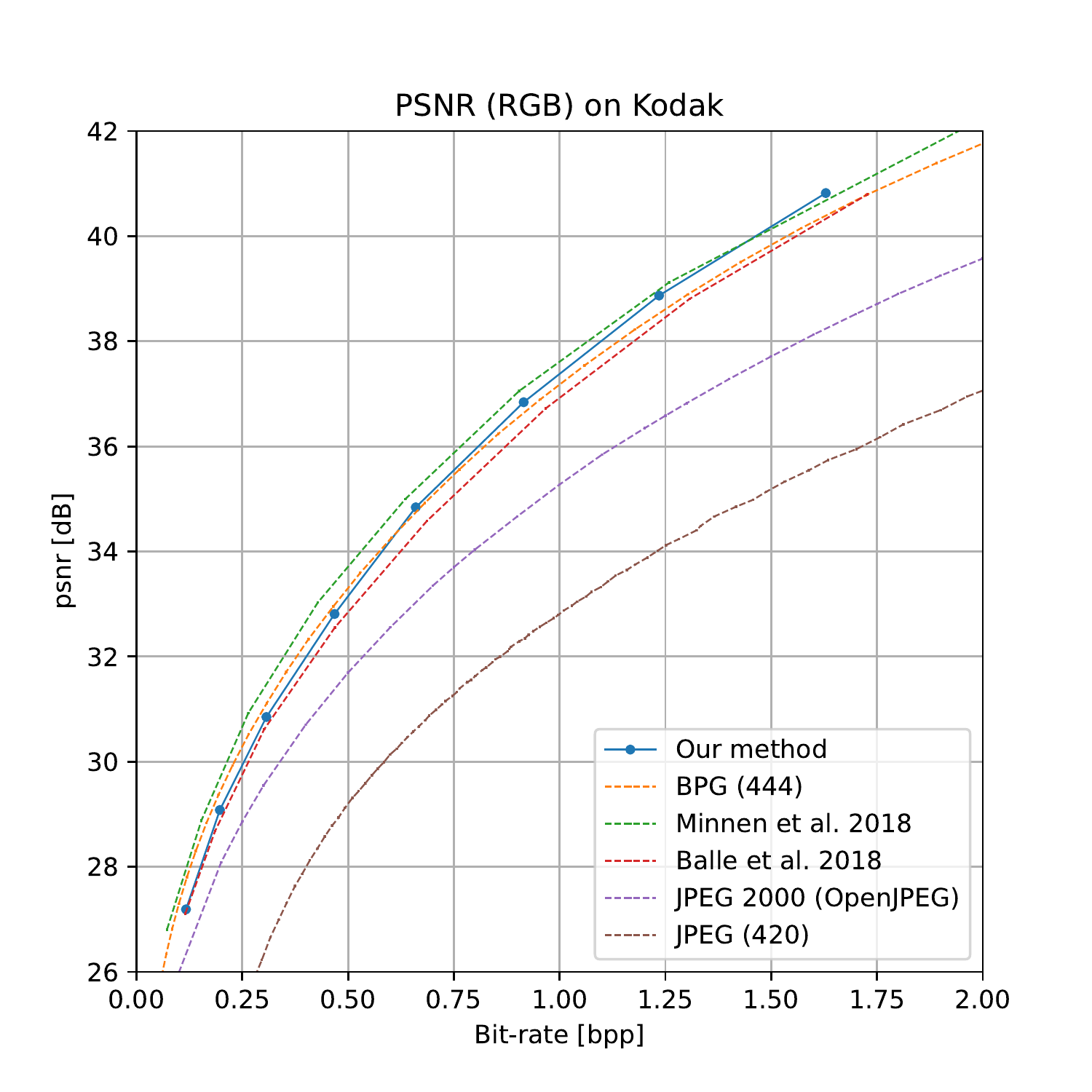} 
\centerline{{\footnotesize (a) Kodak dataset \cite{kodak}}}
\end{minipage}
\hfill
\begin{minipage}{0.49\linewidth}
\includegraphics[trim=20 15 33 35, clip, width=\linewidth]{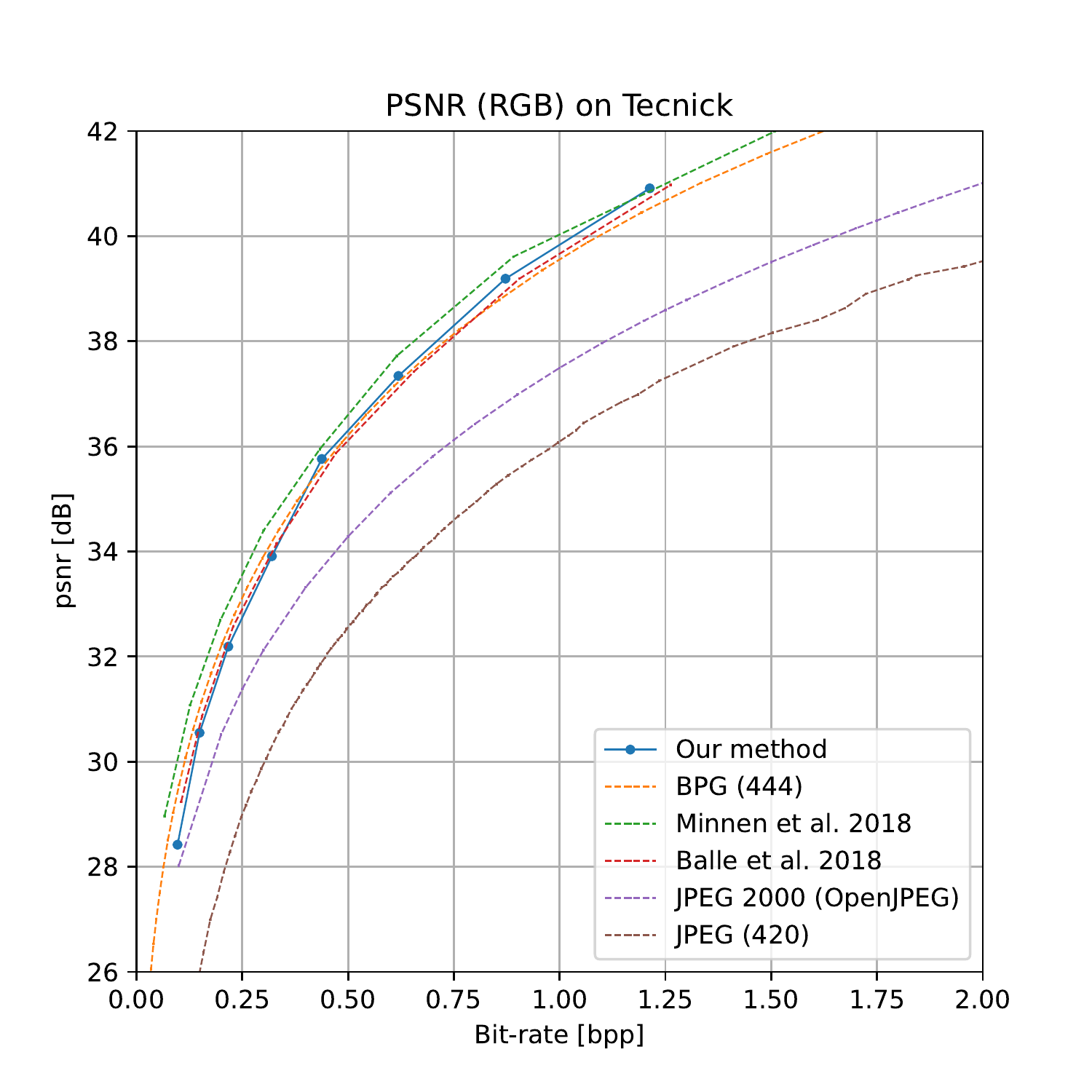}
\centerline{{\footnotesize (b) Tecnick dataset \cite{tecnick}}}
\end{minipage}
\caption{Compression performance results on Kodak and Tecnick datasets. } 
\label{fig:expres}
\end{figure*}

The compression performance results are shown in Figure \ref{fig:expres}. The peak-signal-to-noise-ratio (PSNR) and bits-per-pixel (BPP) values in the figure were calculated by averaging the PSNR and BPP values of all images in each dataset. The compression performance results of compared methods are taken from the publicly available results reported at \cite{tfc} which are prepared by the authors of \cite{balle2018variational, minnen2018joint}.  

\begin{figure*}[tb]
\includegraphics[trim=0 0 0 0, clip, width=0.16\linewidth]{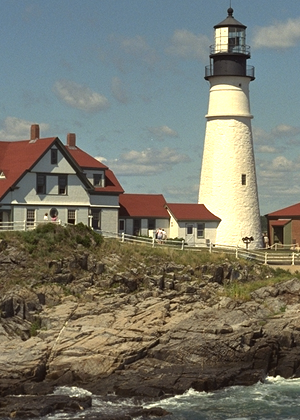} 
\includegraphics[trim=0 0 0 0, clip, width=0.16\linewidth]{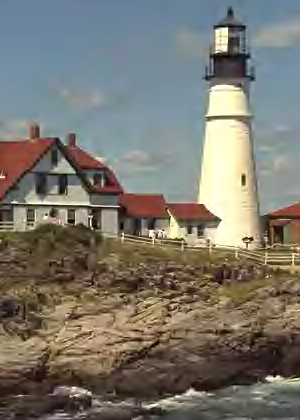} 
\includegraphics[trim=0 0 0 0, clip, width=0.16\linewidth]{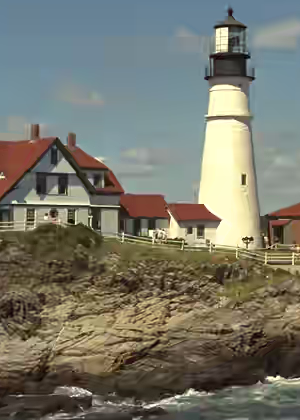} 
\includegraphics[trim=0 0 0 0, clip, width=0.16\linewidth]{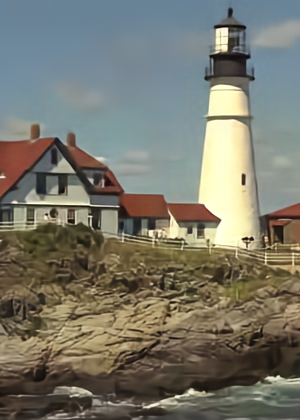} 
\includegraphics[trim=0 0 0 0, clip, width=0.16\linewidth]{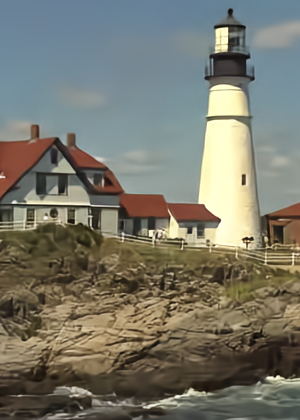} 
\includegraphics[trim=0 0 0 0, clip, width=0.16\linewidth]{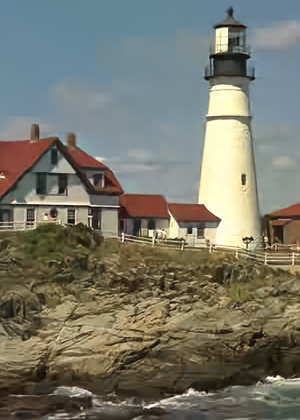} 
\caption{Reconstructed pictures at around 0.22 bpp (Zoom in the pdf file for details.) From left to right with (PSNR, bpp): Original, JPEG2000 (26.84, 0.223), BPG (28.72, 0.222), Balle et al. \cite{balle2018variational} (28.59, 0.224), Minnen et al. \cite{minnen2018joint} (28.96, 0.208), Our method (28.59, 0.223). }
\label{fig:recon_imgs}
\end{figure*}

It can be seen from Figure \ref{fig:expres} that the explored learned block based image compression method provides competitive compression performance. In particular, it provides better compression performance than traditional compression systems JPEG and JPEG2000. Compared to BPG, the explored method provides better compression performance at higher bitrates but slightly inferior performance at lower bitrates. Compared to the learned compression method of \cite{balle2018variational} (Balle et al. 2018), the explored method is on par at the low bitrate end but provides superior results at higher bitrates. Compared to the learned compression method of \cite{minnen2018joint} (Minnen et al. 2018), the explored method provides inferior, though competitive, results at lower bitrates but catches up at the high bitrate end. Some visual results are  provided in Figure \ref{fig:recon_imgs}. More visual results are available at \cite{this-gh}.

There are two aspects which can hinder the performance of the explored learned block based image compression method. The first one is that the explored method is a block based method which processes blocks in a causal order and therefore does not allow (for the compression of a block) fusing information from the right and lower blocks. This is unlike many learned compression methods \cite{balle2018variational, minnen2018joint}. However, as mentioned before, block based compression can also have desirable features, such as easy combination with video compression systems \cite{LEGALL1992129, luthra264} (as discussed in Section \ref{sec:intro}) or internal memory requirement advantages in hardware implementation \cite{hwhevc, hwjp2k}. 

\begin{figure}[tb]
\begin{minipage}{0.49\linewidth}
\includegraphics[trim=3 1 19 30, clip, width=\linewidth]{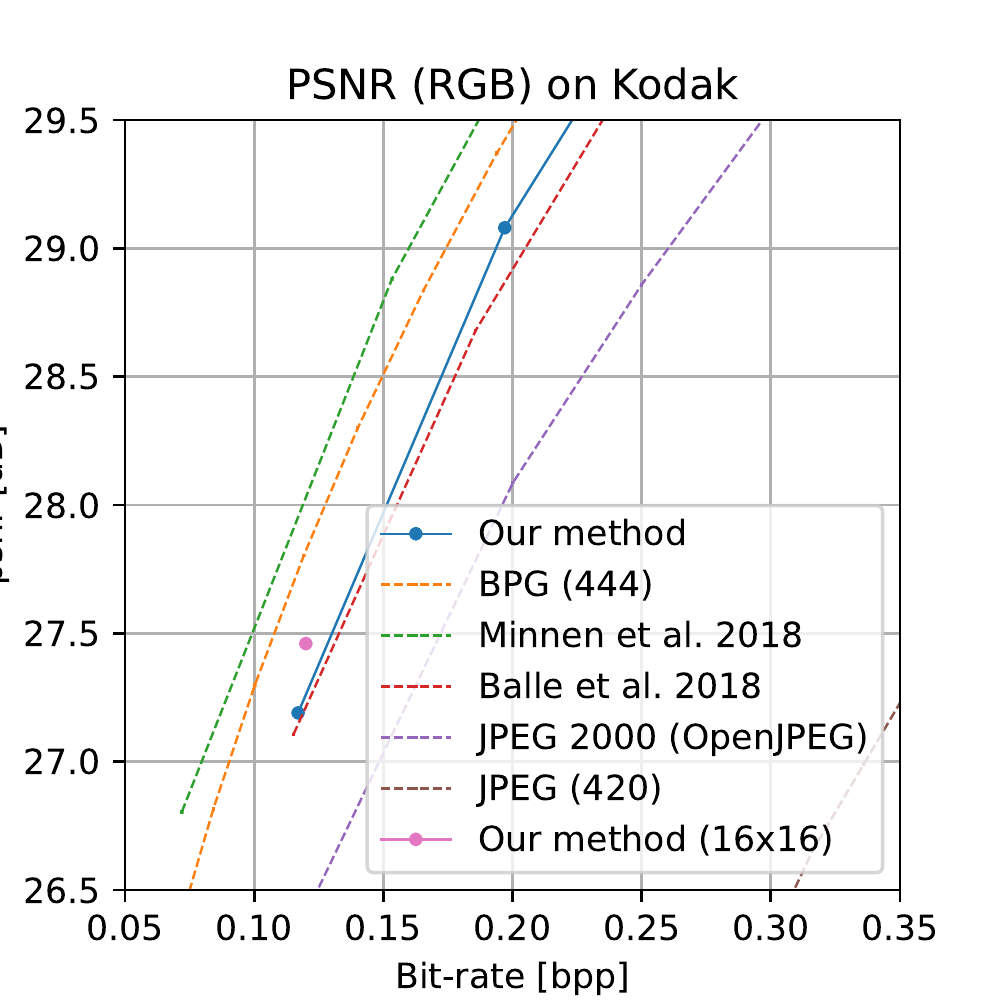} 
\centerline{{\footnotesize (a) 16x16 block size}}
\end{minipage}
\hfill
\begin{minipage}{0.49\linewidth}
\includegraphics[trim=3 1 19 30, clip, width=\linewidth]{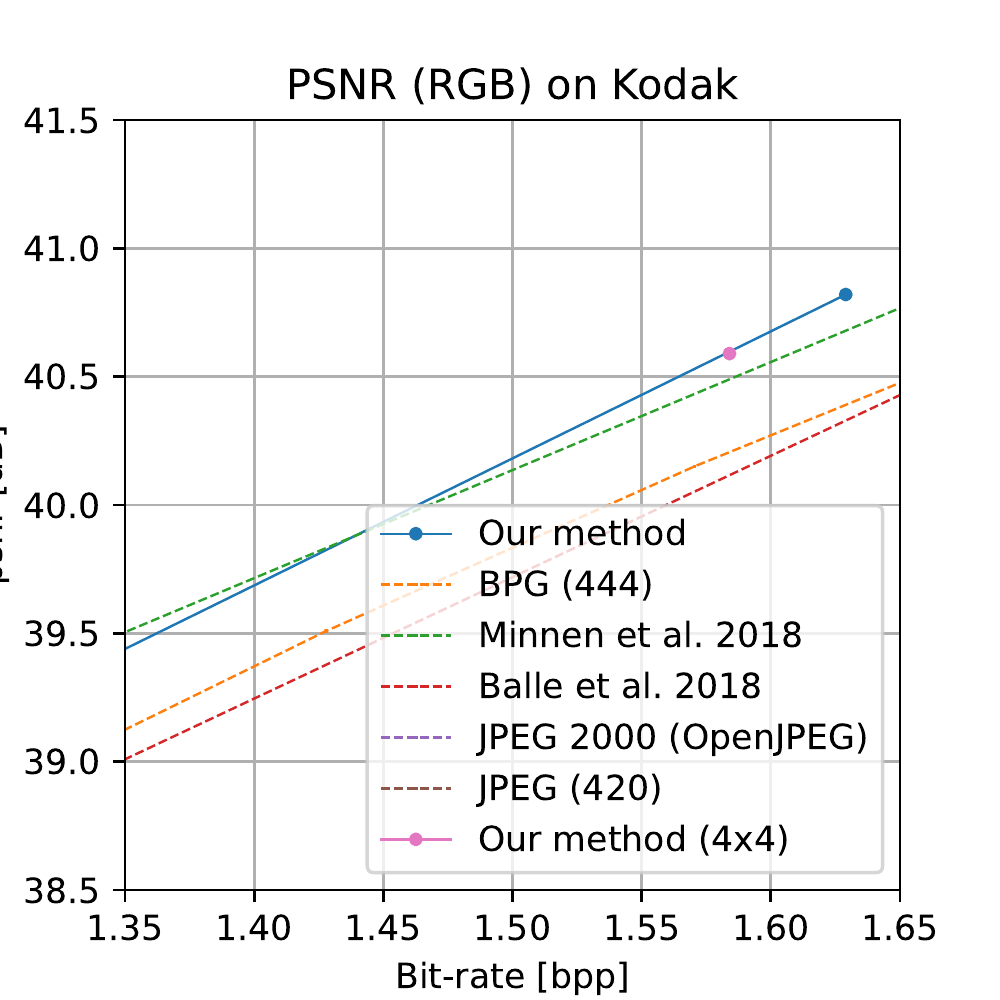}      
\centerline{{\footnotesize (b) 4x4 block size}}
\end{minipage}
\caption{Compression performance results on Kodak dataset with different block sizes of the explored system. }
\label{fig:expres_bxb}
\end{figure}

The second aspect is that the explored system operates at a single block size of 8x8. However, in large smooth regions, the system is likely to provide better compression with a larger block size and in detailed texture regions, it is likely to do better with a smaller block size. To investigate this, the explored system was trained with a (fixed) block size of 16x16 for very low bitrates (i.e. lowest $\lambda$ value) and with a (fixed) block size of 4x4 for high bitrates (i.e. highest $\lambda$ value). The compression results are shown in Figure \ref{fig:expres_bxb}. With a block size of 16x16, the explored system provides slightly better compression at low bitrates. 
With a block size of 4x4, the explored system provides similar (but not better) compression performance at high bitrates. 
These results indicate that supporting multiple block sizes (as traditional compression methods) in the explored learned block based system is likely to increase compression performance, however, it requires careful joint training of multiple block size systems with ACL design and is part of our future research studies. 

\subsection{Computational Complexity Analysis}
\label{ssec:companal}
A computational complexity analysis is provided in Table \ref{tb:cca} using the required number of parameters (of the ANN) and the encoding and decoding times. The encoding and decoding times of our method and those of Balle et al. \cite{balle2018variational} and Minnen et al. \cite{minnen2018joint} (with the implementations in \cite{begaint2020compressai}), all learning based methods, were measured with our computer and run on the GPU\footnote{Nvidia RTX 3060 Ti}, while those of BPG, JPEG and JPEG2000, all traditional methods, were taken from \cite{begaint2020compressai} and run on the CPU. The results in Table \ref{tb:cca} should taken as a rough estimate of computational complexity. A thorough analysis depends on many factors and is out of the scope of this paper.

\begin{table}[tbh]
\begin{small}\begin{center}
\caption{Number of parameters and encoding/decoding times}
\vspace{0.1cm}
{\renewcommand{\arraystretch}{1.2} 
\begin{tabular}{r|cc} 
  & Number of & Run-time\\
  & Parameters & (Enc., Dec.)\\ \cline{1-3}
Our method (HPS1) & 8.6 M & 27.4 , 37.2 s\\ 
Our method (HPS2) & 24.7 M &  29.2 , 41.4 s\\ \hline
Balle et al. 2018 (HPS1) & 4.9 M & 0.06 , 0.03 s\\ 
Balle at al. 2018 (HPS2) & 11.3 M &  0.09 , 0.04 s\\ \hline
Minnen et al. 2018 (HPS1) & 12.6 M & 4.6 , 8.8 s\\ 
Minnen at al. 2018 (HPS2) & 21.8 M &  4.0 , 8.1 s\\ \hline 
BPG & NA & 3.7 , 0.07 s\\ \hline
JPEG2000 & NA &  0.5 , 0.5 s\\ \hline
\end{tabular}
}
\label{tb:cca}
\end{center}\end{small}  
\end{table}

The number of parameters of all learning based methods in Table \ref{tb:cca} are comparable, however, the encoding and decoding times can differ by several orders of magnitude. Traditional methods do not use CNN or other computationally heavy algorithms and thus are typically fast on and optimized for the CPU. Learning based methods require typically orders of magnitude more computations and thus are preferred to run on the GPU. Balle et al.'s method \cite{balle2018variational} has very low encoding and decoding times since it can fully exploit the GPU's parallel computation resources. On the other hand, Minnen et al.'s method \cite{minnen2018joint} can not fully exploit the GPU's parallel computation resources since it requires some serial computations (for the auto-regressive parts of their model) which can not be efficiently computed with standard GPU hardware and software. Similarly, our method requires a block-by-block processing approach, which means CNN computations can not be performed on the entire image at once, which again can not be efficiently computed with standard GPU hardware and software. 

We note that the total number of parameters and therefore the total number computations of Minnen et al.'s method and our method are comparable to those of Balle's method. This indicates that other parallelization methods such as wavefront parallel processing, well-known in traditional video compression \cite{chi2012parallel}, or better optimized GPU hardware/software may enable also faster encoding and decoding times for these methods.

\section{Conclusions}  
\label{sec:conc}
This paper explored an alternative learned block-based image compression approach in which, unlike other block based learned compression approaches in the literature, neither an explicit intra prediction nor a deblocking network neural network is used. A single auto-encoder neural network with block-level masked convolutions is used and the block size is much smaller (8x8). Since the explored system is a closed loop system, a special optimization procedure, the asymptotic closed loop design, was used with standard stochastic gradient descent based training. Experimental results indicate that the explored learned block based approach provides competitive image compression performance. Extending the approach to support multiple block sizes is likely to further increase the compression performance, which is part of our future research directions together with extension to inter-frame compression. 

\bibliography{mybibfile}

\end{document}